\documentclass{optica-article}

\journal{opticajournal}
\usepackage{graphicx}
\usepackage{makecell}
\usepackage{multirow}
\usepackage{booktabs}
\usepackage{appendix}
\usepackage{array}
\usepackage{braket}
\usepackage{amsmath}
\usepackage[marginal]{footmisc}
\usepackage{bm,color,bbm}

\newcommand{\blk}{\color{black}}

\usepackage{float}
\newcolumntype{.}{D{.}{.}{-1}}
\usepackage{amsopn}

\articletype{Research Article}
\usepackage{hyperref}	
\hypersetup{
	colorlinks = true,
	citecolor = {blue},
	linkcolor = {red}
}
\usepackage{lineno}

\begin{document}

\title{Source-independent quantum random number generators with integrated silicon photonics}

\author{Yongqiang Du,\authormark{1,4}  Xin Hua,\authormark{2,3,4}  Zhengeng Zhao, \authormark{1} Xiaoran Sun,\authormark{1} Zhenrong Zhang,\authormark{1} Xi Xiao, \authormark{2,3,$\ast$} Kejin Wei, \authormark{1,$\dagger$} }

\address{\authormark{1} Guangxi Key Laboratory for Relativistic Astrophysics, School of Physical Science and Technology, Guangxi University, Nanning 530004, China\\
	\authormark{2} National Information Optoelectronics Innovation Center (NOEIC), 430074, Wuhan, China\\
	\authormark{3} State Key Laboratory of Optical Communication Technologies and Networks, China Information and Communication Technologies Group Corporation (CICT), 430074, Wuhan, China\\\authormark{4} These authors contributed equally to this paper.
	}



\email{\authormark{$\ast$}xxiao@wri.com.cn} 
\email{\authormark{$\dagger$}kjwei@gxu.edu.cn}


\begin{abstract}
Random numbers play a crucial role in numerous scientific applications. Source-independent quantum random number generators (SI-QRNGs) can offer true randomness by leveraging the fundamental principles of quantum mechanics, eliminating the need for a trusted source. Silicon photonics shows great promise for QRNG due to its benefits in miniaturization, cost-effective device manufacturing, and compatibility with CMOS microelectronics. In this study, we experimentally demonstrate a silicon-based discrete variable SI-QRNG. Using a well-calibrated chip and an optimized parameter strategy, we achieve a record-breaking random number generation rate of 7.9 Mbits/s. Our research paves the way for integrated SI-QRNGs.
\end{abstract}


\section{Introduction}

Random numbers are crucial in daily life and scientific research. Ideally,  a random number generator (RNG) should generate a sequence of numbers that are both uniformly distributed and unpredictable~\cite{2017-Herrero-review}. However, pseduo- or classical RNGs, relying on deterministic computational algorithms or classical physical processes, exhibit predictability and long-range correlation, causing errors in scientific simulations~\cite{1990-pseudorandom}. Such pseudo-randomness can lead to catastrophic errors in scientific simulations, especially in fields like secure communication~\cite{BB84,2020-Xu-review}, lottery, and blockchain technology, where high information security and privacy are essential. In contrast, quantum random numbers~\cite{2016-Antonio,2016-ma-review,2022-Mannalath} arise from the inherent randomness of quantum mechanics, offering random numbers characterized by both uniformity and unpredictability.

Quantum random number generators (QRNGs) have seen decades of development and experimental implementation in various quantum sources. These sources are primarily based on single photons~\cite{2008-Dynes, 2014-Nie}, laser phase fluctuations~\cite{2010-Guo, 2012-Xu, 2020-Wen}, and vacuum states~\cite{2010-Gabriel, 2016-Shi,2019-Zhou}. However, imperfections in QRNG systems can introduce deviations from the ideal model used in security analysis. These deviations can lead to the risk of side information leakage, making the generated random numbers predictable.

Based on the level of trust devices, QRNGs can be classified into three main types.  The first type is fully trusted QRNG which assume both the source and devices are trusted and  can generate high-speed random numbers~\cite{2015-Nie,2021-Guo, 2021-Gehring}. The second type is device-independent QRNGs~\cite{2010-Pironio, 2018-Bierhorst, 2018-Liu, 2020-Zhang, 2021-Shalm}, which generate true random numbers without making any assumptions in the sources and measurement devices. However, this type of QRNG requires loophole-free Bell tests, resulting in low rates in practical applications.  The last type is semi-device-independent QRNG, which requires reasonable assumptions to bound the side information. These assumptions relate to sources ~\cite{2016-Nie-MDI-QRNG, 2021-Mironowicz,2023-Wang}, measurement devices~\cite{2020-Drahi, 2022-Cheng, 2023-Zhang}, the dimension of the underlying Hilbert space~\cite{2015-Lunghi, 2021-Mironowicz } to achieve reasonably fast and secure random number generation. \blk

Source-independent QRNG (SI-QRNG)~\cite{2014Vallone,2016-Cao} is a representative scheme of semi-device-independent QRNGs. The core idea of SI-QRNG is to use measurements for source monitoring, which is often more challenging to characterize than the measurement devices themselves. In such situations, it's necessary to randomly switch between various measurement settings, often using complementary settings, to thwart the source (assumed to be under the control of an adversary) from predicting the next measurement. The feasibility of SI-QRNG has been experimentally demonstrated using bulk optical devices~\cite{2016-Cao, 2019-Li,2017-Marangon}. Substantial efforts have been dedicated to improving performance and practical security~\cite{2018-Avesani,2019-XuBJ,2020-Ma, 2020-Lin,2022-Lin,2023-Liu,2023-Lin}. Similarly to trusted-device QRNGs~\cite{2016-Abellan,2018-Francesco-OE,2018-Francesco, 2019-Thomas,  2020-Acerbi, 2021-Regazzoni,2021-Bai,2023-Bruynsteen}, integrated photonics are employed in SI-QRNG to reduce both size and cost. Remarkably, the random number generation rate reaches Gbps when using continuous-variable states~\cite{2023-Bertapelle}. However, an integrated discrete-variable SI-QRNG has not yet been reported.

In this study, we present a discrete-variable SI-QRNG implemented on a silicon integrated chip. The system features a straightforward structure, mainly comprising a laser diode (LD) chip, a polarization decoder chip with integrated polarization tracking, and single photon detectors (SPDs). By well calibration of chip and optimized parameters based on the protocol in Ref.~\cite{2016-Cao}, our chip QRNG achieved a quantum bit error rate (QBER) of 0.12\% and generated random numbers at a rate of 7.9 Mbps. This sets a new record for discrete-variable SI-QRNG systems (refer to Table~\ref{DV-QRNG_list}). The final random numbers also meet all NIST test suite criteria. The experimental results of this study illustrate the feasibility of implementing SI-QRNG with silicon integrated devices, offering a cost-effective, robust, and scalable integrated QRNG module for next-generation secure communication, the Internet of Things, and scientific simulations.

\begin{table*}[!ht]
	
	\centering 
	\caption{ A list of discrete variable SI-QRNG experiments.}
	\renewcommand
	\arraystretch{1.5}
	\tabcolsep=0.4cm
	\scalebox{0.8}
	{
		\begin{tabular}{cccccc} 
			\hline\hline
			Reference & $f$ (MHz) & $\mu$ & $e_{bX}$ (\%) & Efficiency (\%) & $R$ (bps)
			\\ \hline
			Cao \emph{et al}.~\cite{2016-Cao} & $1$ & $1$ & $2$ & $45$ & $5\times10^3$ 
			\\
			Li \emph{et al}.~\cite{2019-Li} & $4$ & $14.4$ & $0.33$ & $10$ & $1.81\times10^6$
			\\
			Lin \emph{et al}.~\cite{2022-Lin} & $20$ & $10$ & $1$ & $13$ and $17$ & $3.37\times10^6$
			\\
			Liu \emph{et al}.~\cite{2023-Liu} & $5$ & $9.6$ & $3.5$ & $39$ & $5.05\times10^5$
			\\
			Lin \emph{et al}.~\cite{2023-Lin} & $20$ & $10^a$ & N/A & $13$ and $17$ & $1.34\times10^6$
			\\
			This work. & $50$ & $36.58$ & $0.124$ & \makecell{$Z$ basis: $1.76$ and $1.56$\\$X$ basis: $1.79$ and $1.79$} & $7.94\times10^6$
			\\ \hline\hline
			
			\multicolumn{6}{l}{  $^a$The estimated value from the graph of the article.
				N/A: It is not mentioned in the article.  } \\ 
			
			\multicolumn{6}{l}{  Note added: Ref.~\cite{2023-Liu,2023-Lin} has addressed a more comprehensive range of practical security issues. } \\ 
			
		\end{tabular} 
	}	\centering \label{DV-QRNG_list}
	
\end{table*}

\begin{figure*}[!hbt]
	\centering
	\includegraphics[width=0.8\linewidth]{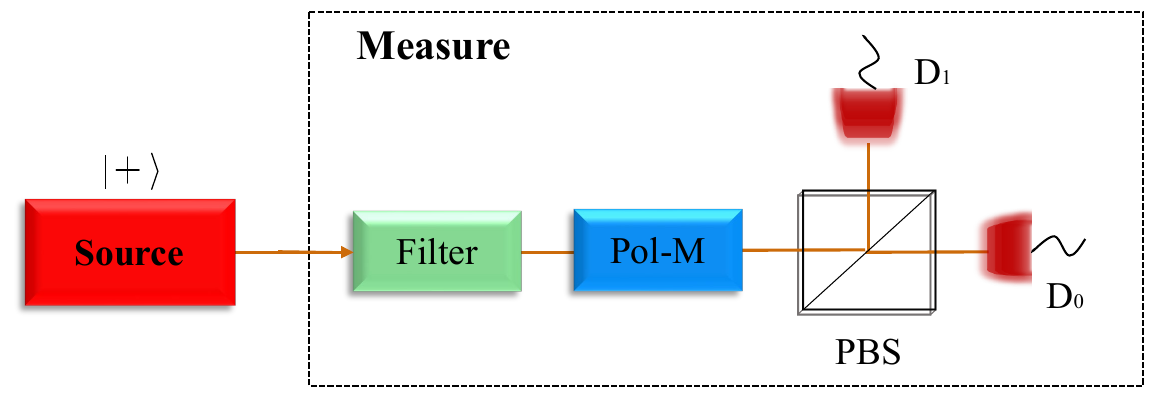}
	\caption{The schematic of a SI-QRNG protocol. It includes an untrusted source and a trusted measurement device. The measurement section consists of components Filter, optical filter; Pol, polarization modulator; PBS, polarized beam-splitter; $D_0$ ($D_1$), the threshold detectors.}
	\label{SI-QRNG}
\end{figure*}

\section{SI-QRNG Protocol}

We first brieftly review the SI-QRNG protocol~\cite{2016-Cao}. A schematic of a discrete variable SI-QRNGs protocol is shown in  Fig.~\ref{SI-QRNG}. In this setup, an untrusted randomness source emits light pulses. Subsequently, a trusted measurement device performs projection measurements on these pulses, which leads to the generation of a random sequence. The operational details of the SI-QRNGs protocol are as follows:

1. \textbf{Source:} An untrusted random source is used as the state preparation device to prepare a polarized state $|+\rangle=(|H\rangle+|V\rangle) / \sqrt{2}$, where $|H\rangle$ and $|V\rangle$ represent horizontal and vertical polarization states, respectively. Because the source is untrusted and may be controlled by eavesdroppers, the quantum state it prepares has arbitrary and unknown dimensions. In this scenario, randomness can still be quantified and extracted through the following steps.

2. \textbf{Squashing:} The squashing process involves mapping quantum states of arbitrary dimensions prepared by the untrusted source into qubits or vacuum states. The vacuum components is considered to account for optical losses and quantum efficiency. In practical experiments, the squashing process is achieved by introducing a series of different types of optical filters to filter out unexpected optical modes.

3. \textbf{Random sampling:} During the measurement phase, Alice randomly chooses to measure the quantum state in either the X $=\{(|H\rangle \pm|V\rangle) / \sqrt{2}\}$ or Z $=\{|H\rangle,|V\rangle\}$ basis based on a random seed. The length of the random seed is exponentially smaller than the extracted random number. In practice, if the total bit number of measured is $N$, where $N_x$ bits are measured in the X basis and $N_z$ bits are measured in the Z basis. It's important to note that this protocol is loss-tolerant. In an ideal scenario, the measurement device would choose the measurement basis after confirming that the received quantum state is not in the vacuum state. However, in experimental settings, the measurement basis is typically chosen before confirming whether the state is in the vacuum state or not, typically by observing whether the detectors register a click. The measurement device cannot predict where the losses will occur. Therefore, the impact of losses only results in a reduction in the sizes of $N_x$ and $N_z$ but the actual positions for effectively measuring the X and Z bases remain random.

4. \textbf{Parameter estimation:} 
When choosing to measure in the X basis, error click events include both the response outcome of $|-\rangle$ and half of the double click events. $\theta$ is statistical deviation~\cite{2010-Fung} estimating the phase error rate of the Z-basis using the bit error rate of the X-basis, given by:
\begin{equation}\label{theta}
	\begin{split}
		\varepsilon_{\theta}=\operatorname{Prob}\left({e}_{{p} Z}>{e}_{{bX}}+\theta\right) \leq \frac{1}{\sqrt{{p}_{{X}}\left(1-{p}_{{X}}\right) {e}_{{bX}}\left(1-{e}_{{bX}}\right) {N}}} 2^{-{N} \xi(\theta)},       	
	\end{split}  
\end{equation} 
where $\xi(\theta)=H(e_{bX}+\theta-p_{X}\theta)-p_{X} H(e_{bX})-(1-p_{X}) H(e_{bX}+\theta)$. $p_{X}$ is the selection probability of the X-basis. $e_{bX}$ is the QBER of the X-basis. $H(x)=-x \log _{2}(x)-(1-x) \log _{2}(1-x)$ is the binary Shannon entropy function.  Due to the no any assumptions about the source, it may emit multiphoton states. When using the threshold detectors, it may lead to the occurrence of double click events, further increasing the error rate $e_{bX}$ in the X basis and reducing the extraction of random bits.

5. \textbf{Randomness extraction:} By employing the Toeplitz matrix hashing method~\cite{1993-Mansour}, we use $N_{Z} H\left(e_{b x}+\theta\right)$ bits to correct phase errors, where the failure probability is $2^{-t_{e}}$~\cite{2011-Ma}. Given all these, the number
of extracted random bits is 
\begin{equation}\label{R}
	\begin{split}
R = N_{z}-N_{z} H\left(e_{bX}+\theta\right)-t_{e},      	
	\end{split}  
\end{equation}
 with the failure probability (in trace-distance
measure) is given  by $\varepsilon =\sqrt{\left ( \varepsilon_\theta + 2^{-t_e}   \right )\left ( 2-\varepsilon _{\theta } - 2^{-t_e} \right )  } $.

\begin{figure*}[!hbt]
	\centering
	\includegraphics[width=0.8\linewidth]{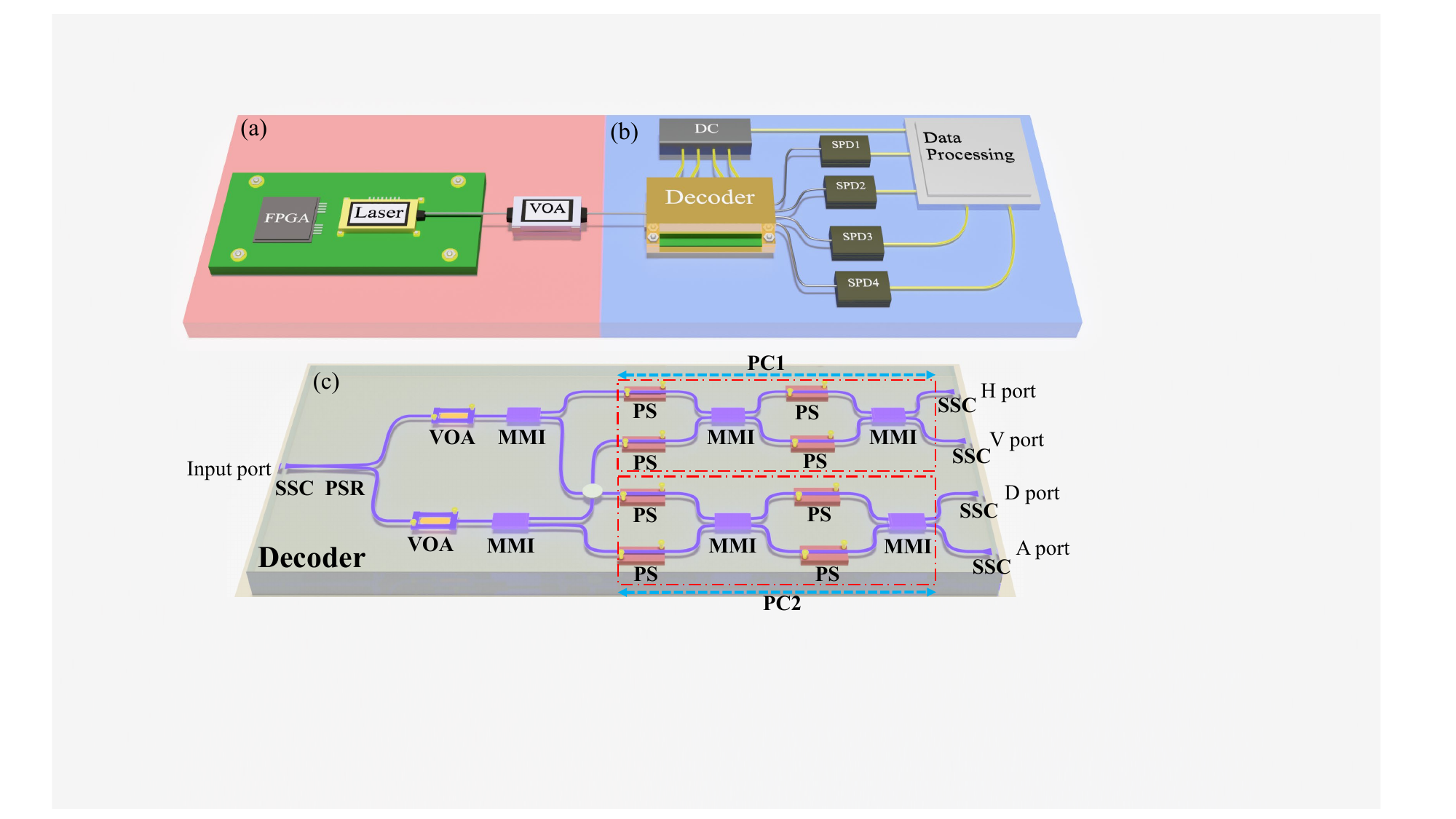}
	\caption{Silicon-based SI-QRNG system: (a) Untrusted randomness light source part. Laser, a laser diode; VOA, a variable optical attenuation. (b) Detection parts. Decoder, polarization state demodulation chip; SPD, single photon detectors; Data processing, including a time digital convert and a personal computer; DC, programmable DC power supply. (c) Schematic of the polarization state demodulation chip. SSC, spot-size converter; PSR, polarization splitter-rotator; VOA, variable optical attenuator; MMI, multimode interferometer; PS, thermo-phase shifter; }
	\label{setup}
\end{figure*}

\section{Setup}

 Figure~\ref{setup} shows the  schematic of our chip-based SI-QRNG. This system primarily consists of two modules: a randomness source which consists of an untrusted laser source, controlled by a field-programmable gate array (FPGA), as shown in Fig.~\ref{setup} (a), and a trusted photonic polarization measurement, including a decoder and four single photon detectors (SPDs, WT-SPD300, Qasky Co. Ltd.), as illustrated in Fig.~\ref{setup} (b).  The source and measurement are interconnected through single-mode fiber links which also acted as the role of a filter. \blk In the source, a laser (LD, WT-LD, Qasky Co. Ltd.) with a repetition rate of 50 MHz is used to emit pulsed light sequences at a central wavelength of 1550 nm and a pulse width of 200 ps. Subsequently, the laser is attenuated by a variable optical attenuation (VOA, DA-100, OZ Optics Ltd.). In the measurement, a polarization decoder chip is employed to decode the polarization state of the photons, and then are detected by an external SPDs and recorded using a data processing module.

The central device in the SI-QRNG system is a polarization decoder chip, which can monitor the source using a conjugate basis. The schematic of the decoder chip is depicted in Fig.~\ref{setup} (c). The chip was fabricated using a standard silicon photonics foundry and has dimensions of $1.6\times1.7 ~\text{mm}^{2}$. It was packaged using chip-on-board assembly, resulting in a total volume of approximately $3.95\times2.19\times0.90 ~\text{cm}^3$.

Any polarization state $\left | \psi \right \rangle_{pol} =\alpha |H\rangle+\beta|V\rangle$, where $\alpha^{2}+\beta^{2}=1$, transmitted from the light source section is initially coupled into the chip through a spot-size converter (SSC). Subsequently, it is converted into path information using a polarization splitter-rotator (PSR)~\cite{2016-Chen}. Two symmetric multi-mode interferometers (MMIs) are used for passive selection of measurement bases in the Z or X basis.

The Z and X basis measurement sections are primarily constructed using two polarization controllers, PC1 and PC2, on the chip. Each polarization controller consists of a pair of phase shifters (PSs) connected to a Mach-Zehnder interferometer driven by another pair of PSs. Additionally, the power settings of the phase shifters (PSs) on the decoder chip, PC1 and PC2, are adjusted to create a POVM with $\left | \psi \right \rangle_{pol}$ as the reference polarization state.

PC2 is defined as the eigenmeasurement basis for the input quantum state $\left | \psi \right \rangle_{pol}$ in the X basis, while PC1 is configured as the complementary measurement basis for the input quantum state $\left | \psi \right \rangle_{pol}$ in the Z basis. This configuration of the decoder chip leads to the measurement of the incident polarization state in the X basis and generates outputs from the D port, with detection events at the A port classified as errors.

Furthermore, when the incident polarization state is measured in the Z basis, it generates raw random bits and outputs them from either the H or V port. Lastly, the photons measured in the Z or X basis are coupled out of the decoder chip through the SSC. Detection events are collected and post-processed using a data processing module. Detailed information about the principles, fabrication process, and parameters of the polarization state decoder chip can be found in Ref.~\cite{2023-Du,2023-Wei}.

\section{Results}\label{}

The initial step involved characterizing the decoder chip. The chip has a total insertion loss of about 4.6 dB. Approximately 3 dB of this loss is attributed to coupling-in and -out, while the remaining 1.6 dB is due to other components and transmission losses on the chip. The phase shifters (PSs) on the chip have a 3 dB bandwidth of approximately 3 kHz and a half-wave voltage of 0.72 V. The polarization extinction ratio of the decoder chip is more than 26 dB. Additional tests show that the probability of selecting the Z basis is 52.83\%. The quantum bit error rate (QBER) for the incident polarization state in the X basis is 0.12\%. The probability of a random collapse to the H port during Z basis measurement is 47.18\%.

Leveraging the system's configuration and the previously mentioned characterization parameters, we conducted SI-QRNG experiments. To maximize the random number generation rate, we recalibrate the SPDs for higher counting rates, albeit at the expense of reduced detection efficiency. The detection efficiency of the detectors at the H, V, D, and A ports is 1.76\%, 1.56\%, 1.79\%, and 1.79\%, with an average maximize counting rate of 10 MHz, respectively. A total number of $N=10^{10}$ pulses is sent, and the average photon number is set to 36.58. The obtained QBER in $X$  $e_{b X}$ is  0.12$\%$, and  the Z-basis single-click rate  $N_{Z}^{s}$  is $1.85\times 10^9$. We then use the security proof in Ref.~\cite{2019-Li}, which takes into account the imperfections of the measurement device, including detector efficiency mismatches and complementary  measurement bases.  The final random number generation rate under a finite regime can be estimated using the following modified formula:
\begin{equation}\label{R_finite}
	\begin{split}
		R_{\text {final }}=\frac{2 \min \left(\eta_{0}, \eta_{1}\right)}{\eta_{0}+\eta_{1}}\left[-2 N_{Z}^{s} \log _{2} \max _{X, Z}\left|\left\langle X\mid Z\right\rangle\right|-N_{z}^{s} H\left(e_{b X}+\theta\right)-t_{e}\right],       	
	\end{split}  
\end{equation}
where $\eta_{0}$ and $\eta_{1}$ are the detection efficiencies of the Z-basis detectors. Term $-2\log _{2} \max _{X, Z}\left|\left\langle X\mid Z\right\rangle\right|$ represents the incompatibility between the Z and X measurement bases. $N_{Z}^{s}$ is the total number of single click events on the Z-basis.

\begin{figure}[h]
	\centering
	\includegraphics[width=0.6\linewidth]{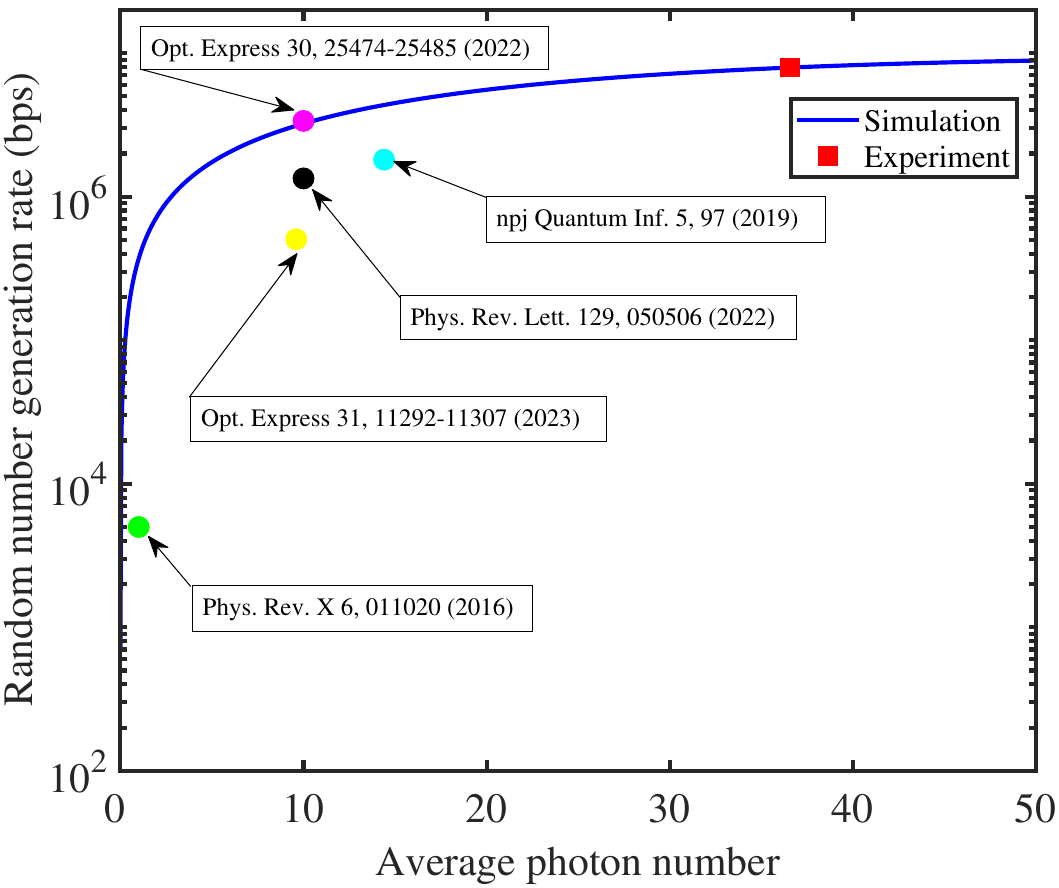}
	\caption{Final quantum random number generation rate with different average photon number. The blue line represent the simulation results based on our experimental system parameters, and the red square represent the experimental results. We also plotted the highest quantum random number generation rates for different SI-QRNGs experiments (circular diagram). }
	\label{Rate}
\end{figure}

 \begin{figure}[!hb]
	\centering
	\includegraphics[width=1\linewidth]{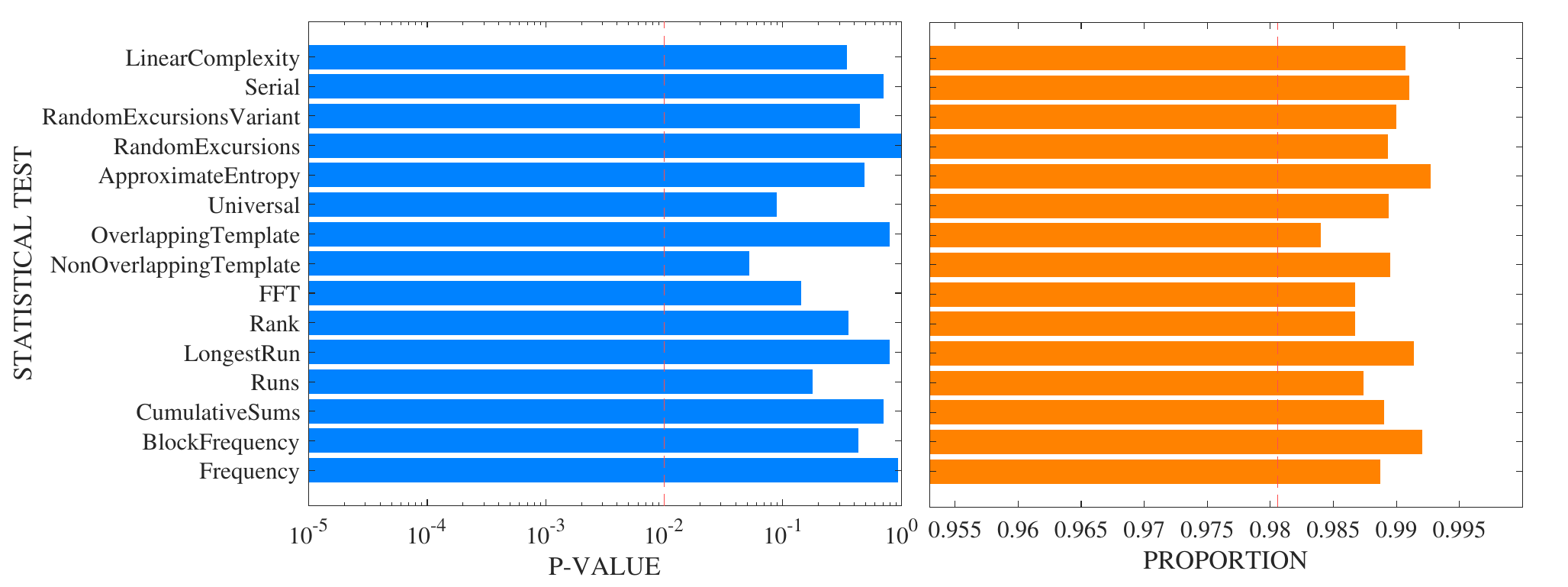}
	\caption{The P-value and proportion value of NIST test terms. We subjected the random numbers collected and processed in the experiment to the standard NIST test suite using 1500 samples of 1 Mbit.  In the case of multi-outcome tests, a further KS uniformity test is applied to the p-values and the corresponding proportions are averaged. \blk }
	\label{NIST_Value}
\end{figure}

Finally, an quantum random number generation rate of  7.94 Mbps is estimated.  The corresponding experimental results are depicted in Fig.~\ref{Rate}, and detailed data is provided in~\ref{rawdate}. Here, the blue line represents a simulated results using the model presented in ~\ref{simulation}. For comparison, we also plotted quantum random number generation rates in  previous  SI-QRNG experiments (represented by circles). Detailed data comparisons can be found in Table~\ref{DV-QRNG_list}. It can be seen that, our work represents the highest rate and In the table, only our work was conducted using silicon-based integrated chips.  It's worth to mention that we employed the same security analysis scheme as that in Ref.~\cite{2019-Li}, making our chip-based QRNG more secure than the original protocol~\cite{2016-Cao}.  Ref.~\cite{2023-Liu,2023-Lin} take into account more comprehensive imperfection factors related to measurement devices.

We then used the Toeplitz Matrix Hashing method~\cite{1993-Mansour} to derive the final random numbers from the raw random bits. To assess the randomness of the generated random numbers, we conducted the NIST SP 800-22 test suite on the final random numbers. The significance level was set at 0.01. A test is considered passed if the P-value for each NIST test term is greater than 0.01, and the corresponding proportion value falls within the confidence interval of $(1-\alpha) \pm 3 \sqrt{(1-\alpha) \alpha / n}=0.99 \pm 0.00944$. Detailed test results from NIST can be found in Fig.~\ref{NIST_Value}. The P-values and proportions for each test passed indicate that the experimentally obtained random numbers exhibit the expected statistical characteristics.

\section{Discussion}

In this study, we propose and experimentally demonstrate a chip-based SI-QRNG that generates random sequences by measuring photon polarization from an untrusted source. Our photon polarization decoding chip provides convenient reconfigurability, allowing polarization state analysis and compensation without requiring additional devices. Furthermore, thanks to the inherent stability and superior polarization extinction ratio of our decoder chip, the SI-QRNG system achieved a low error rate of 0.12\%. The final random number generation rate reaches 7.94 Mbps. We subjected the generated random numbers to NIST testing. Compared to recent discrete-variable SI-QRNGs, our work achieves a balance between integration, speed, and security. Our proposed integrated SI-QRNG offers benefits including affordability, low power consumption, robustness, security, and chip-level integration. It is ideal for portable mobile platforms, airborne applications, and satellite-based QKD applications~\cite{2021-Xue}.

In future research, higher counting rates can be achieved using state-of-the-art high-performance detectors~\cite{2018-Julian,2019-Zhang}, leading to increased quantum random number generation rates. Additionally, integrating single-photon detectors (SPDs) is crucial for advancing the chip-scale integration of SI-QRNG systems. Currently, semiconductor SPDs based on germanium-on-silicon (Ge on Si) technology, employing charge absorption and multiplication regions, offer a promising path for developing future chip-scale SPDs operating at room temperature~\cite{2017-Martinez,2019-vines}. Furthermore, an intriguing approach involves integrating our proposed SI-QRNGs with advanced security protocols~\cite{2023-Lin} to enhance the overall system's security. The integrated SI-QRNG system, based on a standard silicon photonics platform, offers excellent scalability and CMOS compatibility. It can be seamlessly integrated with quantum key distribution (QKD) and central processing unit  systems for random number generation. 

\appendix
\begin{appendices}
 
\renewcommand{\thesection}{Appendix~\Alph{section}}

\section{Simulation}\label{simulation}

In order to characterize the random number generation rate achievable by our chip-based SI-QRNG system at different average photon numbers, we have established a response probability model based on the method described in Ref.~\cite{2019-Fan-yuan}, can be represented as 
\begin{equation}\label{response-probability}
	\begin{split}
p_{0}=1-e^{-\mu p_{Z} \eta_{0}M_{0}^{Z}}\left(1-Y_{0}\right) , p_{1}=1-e^{-\mu p_{Z}\eta_{1}M_{1}^{Z}}\left(1-Y_{0}\right)  , \\   p_{+}=1-e^{-\mu p_{X} \eta_{+}M_{+}^{X}}\left(1-Y_{0}\right) , p_{-}=1-e^{-\mu p_{X} \eta_{-}M_{-}^{Z}}\left(1-Y_{0}\right)  , 	
	\end{split}  
\end{equation} 
where $\mu$ represents the average photon number per pulse from the weak coherent source, $p_{Z} (p_{X})$ stands for the probability of selecting the Z (X) measurement basis, $\eta_{0}$, $\eta_{1}$, $\eta_{+}$, $\eta_{-}$ denote the corresponding detection efficiencies of the SPDs, and $Y_{0}$ represents the dark count rate. $M_{\alpha }^{\beta},\alpha \in \left \{ 0,1,+,- \right \}, \beta \in \left \{ {Z,X} \right \} $ represents the probability of obtaining the measurement outcome $\alpha$ in the $\beta$ measurement basis, since the incident photons are in the eigenstate of the X measurement basis, which we define as $\left | +  \right \rangle $,  theoretically $M_{1}^{Z}$ = $1-M_{0}^{Z}$ = 0.5. $M_{+}^{X}$ represents the intrinsic error rate of the X measurement basis, thus $M_{+}^{X} = 1 -M_{-}^{X}$.

Furthermore, according to Eq.~\ref{response-probability}, the probabilities of single-click events and double-click events in the Z measurement basis are given by 
\begin{equation}\label{theta}
	\begin{split}
& Q_{Z}^{s}=[p_{0}\left(1-p_{1}\right)+p_{1}\left(1-p_{0}\right)](1-p_{+})(1-p_{-}), \\&
 Q_{ Z}^{d}=p_{0} p_{1}(1-p_{+})(1-p_{-}),	
	\end{split}  
\end{equation} 
and the QBER of the X measurement basis can be calculated using 
\begin{equation}\label{theta}
	\begin{split}
		 e_{bX}=\frac{p_{-}\left(1-p_{+}\right)+\frac{1}{2} p_{+}p_{-}}{p_{-}\left(1-p_{+}\right)+p_{+}\left(1-p_{-}\right)+ p_{+}p_{-}}.	
	\end{split}  
\end{equation}

   Other parameters are listed in Table~\ref{QRNG-Simulation-parameter}. The Z-basis single-click count is $N_{Z}^{s} =ftQ_{Z}^{s}$. By combining the aforementioned parameters and using  Eq.~\ref{R_finite}, the variation of quantum random number with the average photon number is obtained. The simulation results are depicted by the blue solid line in Fig.~\ref{Rate}.
 
\begin{table*}[h]
	\centering
	\caption{Simulation parameters.  $f$ is the system repetition rates, $t$ is the total accumulated time, $P_{Z}$ is the probability of
		choosing $Z$ basis, $\eta_{0}$ ($\eta_{1}$, $\eta_{+}$, $\eta_{-}$) is the detection efficiency for $H$ ($V$, $D$, $A$) polarization state, $M_{0}^{Z}$ ($M_{-}^{X}$) is the collapse probability of the incident polarization state when measured using the $Z$ ($X$) basis, $t_e$ is the parameter Alice picks up by balancing the failure probability and the final output length. }
	\renewcommand
	\arraystretch{1.5}
	\tabcolsep=0.4cm
	\scalebox{0.7}
	{\begin{tabular}{cccccccccc}
			\hline \hline
			\multicolumn{1}{c}{$f$ (MHz)}&
			\multicolumn{1}{c}{$t$ (s)}&
			\multicolumn{1}{c}{$P_{Z}(\%)$}&
			\multicolumn{1}{c}{$\eta_{0}$ (\%)}&
			\multicolumn{1}{c}{$\eta_{1}$ (\%)}&
			\multicolumn{1}{c}{$\eta_{+}$ (\%)}&
			\multicolumn{1}{c}{$\eta_{-}$ (\%)}&
			\multicolumn{1}{c}{$M_{0}^{Z}$ (\%)} &
			\multicolumn{1}{c}{$M_{-}^{X}$ (\%)} &
			\multicolumn{1}{c}{$t_e$}
			\\ \hline
			$50$&$200$&$52.83$&$1.76$&$1.56$&$1.79$&$1.79$&$47.18$&$0.12$&$100$\\
			\hline \hline
		\end{tabular}
	}
	\label{QRNG-Simulation-parameter}	
\end{table*}

\section{Detailed experimental results}\label{rawdate}
Table~\ref{table_rawdate} shows the detailed experimental results.

\begin{table*}[!hb]
	\centering 
	\caption{ Experimental parameters and results. $\mu$ is the average photon number of the weak coherent pulse, $t$ is the total accumulated time, $f$ is the system repetition rates, $N_{H}^{s}$ ($N_{V}^{s}$, $N_{D}^{s}$, $N_{A}^{s}$) is the single click counts for the detector of $H$ ($V$, $D$, $A$), $N_{Z}^{d}$ ($N_{X}^{d}$) is the double click counts in $Z$ ($X$) basis, $N_{Z}^{tol}$ ($N_{X}^{tol}$) is the total counts of single and double click in $Z$ ($X$) basis, $e_{bX}$ is the quantum bit error rate of $X$ basis, $\theta$ is the statistical deviation when estimating the phase error rate of Z basis using the bit error rate with X basis, and $R$ is the final quantum random number generation rate. }
	\renewcommand
	\arraystretch{1.5}
	\tabcolsep=0.4cm
	\scalebox{0.7}
	{
		\begin{tabular}{ccccccc} 
			\hline\hline
			$\mu$ & $t$ (s) &$f$ (MHz) & $N_{H}^{s}$ & $N_{V}^{s}$ & $N_{D}^{s}$ & $N_{A}^{s}$ 
			\\ \hline			
			$36.58$ & $200$ & $50$ & $9.29\times10^8$ & $9.23\times10^8$ & $1.92\times10^9$ & $2.02\times10^6$  
			\\      
			$N_{Z}^{d}$ & $N_{X}^{d}$ & $N_{Z}^{tol}$ & $N_{X}^{tol}$ & $e_{bX}$ (\%) & $\theta$ & $R$ (bps)
			\\ \hline
			$1.61\times10^8$ & $7.28\times10^5$ & $2.01\times10^9$ & $1.93\times10^9$ & $0.124$ & $1.23\times10^{-5}$ & $7.94\times10^6$ 
			\\ \hline\hline
		\end{tabular}
	}\centering \label{table_rawdate}
\end{table*}

\end{appendices}

\begin{backmatter}
	\bmsection{Funding} This study was supported by the National Natural Science Foundation of China (Nos. 62171144 and 62031024), the Guangxi Science Foundation (No.2021GXNSFAA220011), and the Open Fund of IPOC (BUPT) (No. IPOC2021A02).
\end{backmatter}

\begin{backmatter}
	\bmsection{Acknowledgments} We thank Shizhuo Li for drawing the diagram of the chip.
\end{backmatter}


\begin{backmatter}
	\bmsection{Disclosures} The authors declare no conflicts of interest.
\end{backmatter}

\begin{backmatter}
	\bmsection{Data Availability} Data underlying the results presented in this paper are not publicly available at this
	time but may be obtained from the authors upon reasonable request.
\end{backmatter}


\bibliography{Chip-SI-QRNG}

\end{document}